\documentclass[a4]{article}

\usepackage{amsthm}
\usepackage{amsmath}
\usepackage{amssymb}
\usepackage{mathrsfs}

\usepackage{lmodern}
\usepackage{color}
\usepackage{dsfont}
\usepackage{graphics}
\usepackage{graphicx}
\newtheorem{Definition}{Definition}

\begin{document}
\title{PRIVACY PRESERVING DISTRIBUTED PROFILE MATCHING IN MOBILE
SOCIAL NETWORK}
\author{\textbf{Rachid CHERGUI} \thanks{%
The research is partially supported by ATN laboratory of USTHB University.%
} \\
{\small USTHB, Faculty of mathematics }\\
[-0.8ex] {\small P.B. 32, El Alia, 16111, Bab Ezzouar, Algeria}\\
[-0.8ex] {\small \texttt{rchergui@usthb.dz 
}}
}
\date{ {\small }}
\maketitle

\begin{abstract}
In this document, a privacy-preserving distributed profile matching protocol is proposed in a particular network context called \emph{mobile social network}. Such networks are often deployed in more or less hostile environments, requiring rigorous security mechanisms. In the same time, energy and computational resources are limited as these heterogeneous networks are frequently constituted by wireless components like tablets or mobile phones. This is why a new encryption algorithm having an high level of security while preserving resources is proposed in this paper. The approach is based on elliptic curve cryptography, more specifically on an almost completely homomorphic cryptosystem over a supersingular elliptic curve, leading to a secure and efficient preservation of privacy in distributed profile matching.
\end{abstract}


\section{Introduction}
Social networking websites, like Facebook~\cite{facebook} with its 900 million active users or Google+~\cite{googlePlus}, are of widespread use in our connected and globalized world. A major trend of these social networks is to attempt to provide instant and real-time access to for users, whatever their location and the connected device they use. This sensible demand from users has led to the development of mobile social networking (MSN) software like Foursquare~\cite{foursquare} and Gowalla~\cite{gowalla}, in which individuals with similar interests are connected together and converse with one another through either tablets or mobile phone. In that approach, mobile apps use existing social networks to create native communities and promote discovery, leading to an improvement of web-based social networks using mobile features and accessibility.
Making new connections according to personal preferences is a crucial service in MSN, where the initiating user can find matching users within physical proximity of him/her. In existing systems for such services, usually all the users directly publish their complete profiles for others to search.  However, in many applications, the users'  personal profiles may contain sensitive information that they do not want to make public. Authors of~\cite{Li11} have presented FindU, a first privacy-preserving personal profile matching scheme, designed for mobile social networks.  In FindU, an initiating user can find from a group of users the one whose profile best matches with his/her; to limit the risk of privacy exposure, only necessary and minimal information about the private attributes of the participating users is exchanged. They speak about a Blind and Permute (BP) protocol. Several increasing levels of user privacy are defined, with decreasing amounts of exchanged profile information.
Authors of this document propose to use a different encryption scheme into the BP algorithm. This new scheme can provide a similar level of security while reducing drastically the computation and communication costs, which is critical in the MSN context. In BP algorithm, encryption over ciphertexts is required. The original method proposed in~\cite{Li11} achieves this requirement using a cryptosystem~\cite{Paillier99} that needs a lot of resources, which is quite incompatible with the constraints related to MSNs. Contrarily, the scheme proposed here is based on elliptic curve cryptography~\cite{boneh}, which leads to smaller keys and cryptograms, low cost computations and shorter communication messages, reducing largely by doing so the batteries consumptions.
The remainder of this document is organized as follows. In Section~\ref{sec:relatedWorks}, related works in the field of privacy-preserving profile matching are proposed. Then, in Section ~\ref{sec:FindU}, we give recall the FindU protocol with related definitions. We give the protocol BP in Section ~\ref{sec:BP}. We construct the homomorphism encryption in Section ~\ref{sec:HE} and we use it in Section ~\ref{sec:mBP} with performance analysis in Section ~\ref{sec:PA}. Section ~\ref{sec:conclusion} conclude this work.
\section{Related Works}
\label{sec:relatedWorks}
The methods used in the field of privacy-preserving distributed profile matching are usually classified into three main categories according to the cryptographic tools they use.
In protocols based on \emph{oblivious polynomial evaluation}, client and a server compute the intersection of the sets corresponding to their profiles, such that the client gets the result while server learns nothing. Homomorphic encryption that allows operations over cipher texts is used to evaluate a polynomial that represents clienta€™s input obviously. This method has been originally proposed in \cite{Freedman04}, through the FNP scheme. Other examples lying in the same category can be found, for instance, in \cite{Kissner0} and \cite{Ye08}. These methods are however impracticable in MSNs because they do not achieve linear computational complexity.
Protocols  based  on \emph{oblivious pseudorandom functions} consist of two parties that securely compute a pseudorandom function, where one of them holds the key while the other
provides the input (set elements). The objective is a secure set intersection. Suppose two parties with private sets wish to learn the intersection set without revealing anything else. Let $P_{1}$ and $P_{2}$ be two parties that have input $X$ and $Y$ respectively and $F$ a pseudorandom function, while $k$ is a key for $F$ belonging to  $P_{1}$. $P_{2}$ compute $\{F_{k}(y) \}_{y \in Y}$ and $P_{1}$ compute $\{F_{k}(x) \}_{x \in X}$ and send the results to $P_{2}$. Thus, $P_{2}$ compare which elements appear in both sets to learn the intersection \cite{Hazay}. The complexity of this method is smaller than the first.
The last category consists of protocols based on so-called commutative encryption.  An encryption scheme $E_{k} ( · ) $ is said to have the commutative property when, for all keys $ k_1 $ and $ k_2 $, we have: $E_{k_1}(E_{k_2}(x)) = E_{k_2}(E_{k_1}(x)) $. For instance, the well known RSA encryption scheme has this commutative property. The main idea when considering privacy-preserving profile matching is thus to use the commutative encryption as a keyed one-way hash function, to generate a mapping for each element $x$ such that no party knows the key \cite{Agrawa}. A commonly related disadvantage of this method is that it often provide a weaker security\cite{Li11}.
Authors of \cite{Li11} have presented a privacy-preserving profile matching called FindU. FindU is a symmetric protocol , i.e., the output is known at the same time by all parties. The characteristics of this scheme is further detailed in the next section.
\section{The FindU Protocol}
\label{sec:FindU}
\subsection{Problem Definition}
In mobile social networks, devices are wirelessly connected (using wireless interfaces such as bluetooth or wifi), thus resources are limited and a certain level of security is required. Authors of FindU algorithm suppose that the connexion is established under public key cryptosystem, where keys are distributed over parties securely. Then, when a party launches a matching, BP algorithm assure sharing a secret securely. Let us define these stages more precisely.
The system consists of $N$ users (parties) denoted as $P_{1},...,P_{N}$, each possessing a portable device. We denote the initiation party \emph{(initiator)} as $P_{1}$. $P_{1}$ launches the matching process and its goal is to find one party that best \emph{matches} with it, from the rest of the parties $P_{2},...,P_{N}$ that are called \emph{candidates}. Each party $P_{i}$'s profile consists of a set of attributes $S_{i}$, which can be strings up to a certain length. $P_{1}$ defines a matching query to be a subset of $S_{1}$ (in the following we use $S_{1}$ to denote the query set unless specified). Also, we denote $n=|S_{1}|$ and $m=|S_{i}|,i>1$, assuming that each candidate has the same set length for the sake of simplicity.
Let us now introduce the following definitions.
\begin{Definition}
The \emph{match} of the set $S_i, i \in \{2, …, N\}$, is by definition the cardinality of $S_{1}\bigcap S_{i}$.
\end{Definition}
\begin{Definition}
The \textit{best match} $P_{i^{*}}$ is defined as the party having the maximum intersection set size with $P_{1}$.
\end{Definition}
$P_{1}$ will first find out $P_{i^{*}}$ via the proposed protocol. Then they will decide whether to connect with other based on their actual intersection set.
\subsection{Adversary Models}
If a party obtains one or more (partial or full) attribute sets without the explicit consents from these users, we said he has achieved an \emph{user profiling}.
In that context, the two following levels of security can be defined \cite{Li11}.
\begin{itemize}
\item \textbf{Honest-but-Curious (HBC) adversary.} In this model, the attacker tries to learn more information than what is allowed, by inferring from the results while honestly following the protocol.
\item \textbf{Malicious adversary.} The attacker tries here to learn more information than allowed by deviating from the protocol run.
\end{itemize}
\subsection{Design Goals}
Here we intend to develop the design goals of FindU scheme. One of the main goals is to defend against profiling attack defined in the previous section. We let the user choose his level of security requirement that we discuss in the next section. By definition, the party $P_{1}$ search among all parties the best that match with him, and at the end, the output of the algorithm will contain the intersection set  between his set query at the profile set of all other parties. By launching FindU, and adversary may obtains all those informations. Thus, we let the user choose his privacy level.
The main security goal is to thwart user profiling attack. Since the users may have different privacy requirement, and as it takes different amount of effort in protocol run to achieve them, we hereby define three levels of privacy where a higher level leaks less information to the adversary. Note that, by default, all of the following include letting $P_{1}$ and the best match $P_{i^{*}}$ learn the intersection set between them at the end of a protocol run.
\begin{itemize}
\item \textbf{Privacy level 1 (PL-1).} When the protocol ends, $P_{1}$ and each candidate $P_{i},1<i\leq N$, mutually learn the intersection set between them, that is, $I_{1,i}=S_{1}\cap S_{i}$. An adversary $A$ should learn nothing beyond what can be derived from the above outputs and private inputs. \newline
If we assume the adversary has unbounded computing power, PL-1 actually corresponds to unconditional security for all the parties under the HBC model . Obviously, in PL-1, $P_{1}$ can obtain all candidates' intersection sets just in one protocol run, thus it reveals too much user information to the attacker, if he assume the role of $P_{1}$. \newline
Therefore we define privacy level 2 in the following.
\item \textbf{Privacy level 2 (PL-2).} When the protocol ends, $P_{1}$ and each candidate $P_{i},1<i\leq N$, mutually learn the size of their intersection set: $m_{1,i}=|S_{1}\cap S_{i}|$. In addition, the best match $P_{i^{*}}$ is allowed to know $m_{1,i}$ values of other $P_{i}s$. The adversary $A$ should learn nothing beyond what can be derived from the above outputs and its private inputs.
\item \textbf{Privacy level 3 (PL-3).} At the end of the protocol, $P_{1}$ and each $P_{i}$ should only learn the ranks of each value $m_{1,i},1<i\leq N$. $A$ should learn nothing more than what can be derived from the outputs and its private inputs. \newline
In PL-3, we can require that $P_{1}$ only contacts the best match $P_{i^{*}}$, such that it only obtains the intersection set $I_{1,i}$ with the best match. In this way, $A$ will need at least $N-1$ protocol runs to know all other user's exact information, such that $A$'s profiling capability is much limited
\end{itemize}
Authors of FindU suggest that the protocol should be \emph{lightweight and practical}, i.e., being enough efficient in computation and communication to be used in MSN. This is why we suggest to introduce homomorphism encryption into the FindU protocol. Readers are referred to \cite{Li11} for a complete decryption of FindU. In order to achieve PL-2, authors introduce homomorphism encryption over cypher-text. For our part, to reduce largely the energy consumption, we suggest to use elliptic curve based encryption.
The Blind and Permute Protocol (BP), part of the FindU system, is presented in the next section, whereas the proposed improvement is detailed in Section~\ref{section:homomorphism}.
\section{Blind and Permute Protocol (BP)}
\label{sec:BP}
The input to BP protocol is a sequence $S=(s_{1},...,s_{n})$ of integer values that is componentwise additively split between $A$ who has $S'=(s'_{1},...,s'_{n})$ and $B$ who has $S''=(s''_{1},...,s''_{n}), such that S=S'+S''$ \cite{Paillier99}, where $+$ stands for the vectorial addition of integers. The output is a sequence $\hat{S}$ obtained from $S$ by:
\begin{enumerate}
\item permuting the entries of $S$ according to a random permutation $\pi$ that is known to neither $A$ nor $B$,
\item modifying the additive split of the entries of $S$ so that neither $A$ nor $B$ can use their share of it to gain any information about $\pi$. We seek a protocol that does this in linear computation and communication complexity.
\end{enumerate}
Observe that it suffices to give a protocol that does half of the job: It blinds and permutes for $A$ according to a random permutation chosen by $B$. Then we can use such protocol a second time with the roles $A$ and $B$ reversed, resulting in a permutation that is the composition of two random permutations: one chosen by $B$ and unknown to $A$, another chosen by $A$ and unknown to $B$. The protocol where $B$ chooses the permutation is given next.
\begin{enumerate}
\item $A$ computes and sends $E_{A}(s'_{1}),...,E_{A}(s'_{n})$ to $B$ (here $E$ is the cryptosystem defined in \cite{Paillier99} whose performance is compared to our scheme in section ~\ref{sec:PA}).
\item$B$ selects $n$ random numbers $r_{1},...,r_{n}$, and for every $i \in {1,...,n}$ he computes $E_{A}(-r_{i})$ and multiplies it by the $E_{A}(s'_{i})$ he received in the first step, thereby obtaining $E_{A}(s'_{i}-r_{i})$.
\item$B$ generates a random permutation $\pi_{B}$ and applies it to the sequence of $E_{A}(s'_{i}-r_{i})$'s computed in the previous step, obtaining a sequence of the form $E_{A}(v'_{1}),...,E_{A}(v'_{n})$ that he sends to $A$. He also applies $\pi_{B}$ to the sequence $s''_{1}+r_{1},...,s''_{n}+r_{n}$, obtaining a sequence $v''_{1},...,v''_{n}$. Note that the sequence  $v'_{1}+v''_{1},...,v'_{n}+v''_{n}$ is a permuted version of $S$ (permuted according to $\pi_{B}$).
\item$A$ decrypts the $n$ items $E_{A}(v'_{1}),...,E_{A}(v'_{n})$ received from $B$, obtaining the sequence $v'_{1},...,v'_{n}$.
\end{enumerate}
In the FindU algorithm (advanced version), BP permit achieving PL-2 level of security.
\section{Homomorphism Encryption}
\label{sec:HE}
\label{section:homomorphism}
We use elliptic curves based cryptography to construct homomorphism encryption function.
\subsection{Operation over Elliptic Curves}
\subsubsection{Addition and Multiplication}
Elliptic curve cryptography (ECC) is an approach to public-key cryptography based on the algebraic structure of elliptic curve over finite fields~\cite{guyeuxVictoria}. Elliptic curves used in cryptography are typically defined over two types of finite fields: prime fields $\mathds{F}_{p}$, where $p$ is a large prime number, and binary extension fields $\mathds{F}_{2^{m}}$ \cite{Cheung05}. In our paper, we focus on elliptic curves over $\mathds{F}_{p}$. Let $p>3$, then an elliptic curve over $\mathds{F}_{p}$ is defined by cubic equation $y^{2}=x^{3}+ax+b$ as the set
$$ \Sigma =\{(x,y)\in \mathds{F}_{p} \times \mathds{F}_{p} \mid y^{2} \equiv x^{3}+ax+b~(mod~ p)\} $$
where $a,b \in \mathds{F}_{p}$ are constants such that $4a^{3}+27b^{2} \neq 0 ~(mod~p)$. An elliptic curve over $\mathds{F}_{p}$ consists of the set of all pairs of affine coordinates $(x,y)$ for $x,y \in \mathds{F}_{p}$ that satisfy an equation of the above form and an infinity point $O$.
The point addition and its special case,  point doubling over $\Sigma$, is defined as follows (the arithmetic operations are defined in $\mathds{F}_{p}$ \cite{ECC}).
Let $P=(x_{1},y_{1})$ and $Q=(x_{2},y_{2})$ be two points of $\Sigma$. Then:
$$
P+Q=\left\{\begin{array}{ll}
O & \mbox{if $x_{2}=x_{1}$ and $y_{2}=-y_{1}$,}   \\
(x_{3},y_{3}) & \mbox{otherwise.}\end{array}\right. $$
where:
\begin{itemize}
\item $x_{3}=\lambda^{2}-x_{1}-x_{2}$,
\item $y_{3}=\lambda\times(x_{1}-x_{3})-y_{1}$,
\item $
\lambda=\left\{\begin{array}{rl}
(y_{2}-y_{1}) \times (x_{2}-x_{1})^{-1} & \mbox{if $P \neq Q$,}   \\
(3x_{1}^{2}+a)\times(2y_{1}^{-1} & \mbox{if $P=Q$.}\end{array}\right. $
\end{itemize}
Finally, we define $P+Q=O+P=P,\forall P \in \Sigma$, which leads to an abelian group $(\sigma,+)$. The multiplication $n\times P$ means $P+P+...+P$ $n$ times, and $-P$ is the symmetric of $P$ for the group law $+$ defined above, for all $P\in \Sigma$.
\subsubsection{Public/Private Keys Generation with ECC}
In this section we show how we can generate the public and private keys for encryption, following the cryptosystem proposed by Boneh et al. \cite{boneh}.
Let $t>0$ be an integer called ``security parameter''. To generate public and private keys, first of all, two $t-bits$ prime numbers must be computed. Therefore, a cryptographic pseudorandom generator can be used to obtain two vectors of $t$ bits, $q_{1}$ and $q_{2}$. Then, a Miller-Rabin test can be applied for testing the primality or not of $q_{1}$ and $q_{2}$. We denote by $n$ the product of $q_{1}$ and $q_{2}$, $n=q_{1}\times q_{2}$, and by $l$ the smallest positive integer such that $p=l\times n-1$. $l$ is a prime number while $p=2~(mod~~3)$.
In order to find the private and public keys, we define a group $H$, which presents the points of the super-singular elliptic curve $y^{2}=x^{3}+1$ defined over $\mathds{F}_{p}$. It consists of $p+1=n\times l$ points, and thus has a subgroup of order $n$, we call it $G$. In another step, we compute $g$ and $u$ as two generators of $G$ and $h=q_{2}\times u$. Then, following \cite{ECC}, the public key will be presented by $(n,G,g,h)$ and the private key by $q_{1}$.
\subsubsection{Encryption and Decryption}
After the private/public keys generation, we proceed now to the encryption and decryption phases:
\begin{itemize}
\item Encryption: Assuming that our message space consists of integers in the set ${0,1,...,T}$, where $T<q_{2}$, and $m$ the (integer) message to encrypt. First, a random positive integer is picked from te interval $[0,n-1]$. Then, the cypher-text is defined by
$$C=m\times g+r\times h \in G,$$
in which $+$ and $\times $ refer to the additive and multiplication laws defined previously.
\item Decryption: once the message $C$ arrived to destination, to decrypt it, we use the private key $q_{1}$ and the discrete logarithm of base $q_{1}\times g$ as follows:
$$m=log_{q_{1} \times g}q_{1}\times C $$
\end{itemize}
\subsection{Homomorphic Properties}
As we have mentioned before, our approach ensures easy encryption/decryption without any need of extra resources. This will be proved in the next section. Moreover, our approach supports homomorphic properties, which gives us the ability to execute operations on values even though they have been encrypted. Indeed, it allows N additions and one multiplication directly on cryptograms. As the product operation will not be used in the profile matching, we will not detail it in this section
Addition aver cypher-texts are done as follows: let $m_{1}$ and $m_{2}$ be two messages and $C_{1},C_{2}$ their cypher-text respectively. Then the sum of $C_{1}$ and $C_{2}$, let call $C$, is represented by $C=C_{1}+C_{2}+r\times h$ where $r$ is an integer randomly chosen in $[0,n-1]$ and $h=q_{2}\times u$ as presented in the previous section. This sum operation guarantees that the decryption value of $C$ is the sum $m_{1}+m_{2}$.
\section{The modified version of BP Protocol}
\label{sec:mBP}
We rewrite the protocol BP with our novel cryptosystem with $E$ meaning the novel algorithm.
\begin{enumerate}
\item $A$ computes and sends $E_{A}(s'_{1}),...,E_{A}(s'_{n})$ to $B$.
\item$B$ selects $n$ random numbers $r_{1},...,r_{n}$, and for every $i \in {1,...,n}$ he computes $E_{A}(-r_{i})$ and add it with the $E_{A}(s'_{i})$ he received in the first step, thereby obtaining $E_{A}(s'_{i}-r_{i})$.
\item$B$ generates a random permutation $\pi_{B}$ and applies it to the sequence of $E_{A}(s'_{i}-r_{i})$'s computed in the previous step, obtaining a sequence of the form $E_{A}(v'_{1}),...,E_{A}(v'_{n})$ that he sends to $A$. He also applies $\pi_{B}$ to the sequence $s''_{1}+r_{1},...,s''_{n}+r_{n}$, obtaining a sequence $v''_{1},...,v''_{n}$. Note that the sequence  $v'_{1}+v''_{1},...,v'_{n}+v''_{n}$ is a permuted version of $S$ (permuted according to $\pi_{B}$).
\item$A$ decrypts the $n$ items $E_{A}(v'_{1}),...,E_{A}(v'_{n})$ received from $B$, obtaining the sequence $v'_{1},...,v'_{n}$
\end{enumerate}
\section{Performance Analysis}
\label{sec:PA}
The experimental results presented in \cite{guyeuxVictoria} compare the performance comparison between RSA and ECC. For the same level of security, say level one, a device operating over RSA need a key of 472 bits while over ECC we need only a key of 46 bits. In \cite{Paillier99}, authors give a performance analysis between a cryptosystem based on  Composite Degree Residuosity Classes CDRC, which is the scheme that is proposed in the BP algorithm. First, RSA is better then CDRC in term of computational complexity. CDRC offer a security level equivalent to $Class[n]$ while RSA is equivalent to $RSA[n,\mathds{F}_{4}]$ and we have \cite{Paillier99} $$ RSA[n,\mathds{F}_{4}] \Rightarrow Class[n] $$
On the other hand, for the same key size, CDRC require 5120 elementary operations for encryption while RSA need only 17 operations.
All those results prove the efficiency of ECC in term of performance.
\section{Conclusion and Future Work}
\label{sec:conclusion}
An homomorphic encryption scheme that enhances the performance of the FindU algorithm has been proposed in this document.
Achieving the PL-3 security level is the main open problem not yet resolved. In future work, homomorphic encryption will be investigated in order to solve this issue.

\end{document}